# Single-hole GPR reflection imaging of solute transport in a granitic aquifer

Caroline Dorn[1], Niklas Linde[1], Tanguy Le Borgne[2], Olivier Bour[2], Ludovic Baron[1]

[1]Institute of Geophysics, University of Lausanne, Lausanne, Switzerland;

[2]Geosciences Rennes, UMR CNRS 6118, University of Rennes 1, Rennes, France.

**Abstract**

Identifying transport pathways in fractured rock is extremely challenging as flow is often organized in a few fractures that occupy a very small portion of the rock volume. We demonstrate that saline tracer experiments combined with single-hole ground penetrating radar (GPR) reflection imaging can be used to monitor saline tracer movement within mm-aperture fractures. A dipole tracer test was performed in a granitic aquifer by injecting a saline solution in a known fracture, while repeatedly acquiring single-hole GPR sections in the pumping borehole located 6 m away. The final depth-migrated difference sections make it possible to identify consistent temporal changes over a 30 m depth interval at locations corresponding to fractures previously imaged in GPR sections acquired under natural flow and tracer-free conditions. The experiment allows determining the dominant flow paths of the injected tracer and the velocity (0.4-0.7 m/min) of the tracer front.

**1. Introduction**

Identifying and characterizing individual permeable fractures and corresponding flow paths at the local field-scale (1-100 m) is an important, but largely unresolved problem with implications for designing waste disposals (nuclear, toxic waste, $CO_2$) and extraction of natural resources (oil, gas, heat, water). Hydrogeological investigations of fractured rock are commonly based either on local measurements in the vicinity of boreholes or hydraulic or tracer inference testing that provides low-resolution integrated information between boreholes or packed-off borehole intervals. Geophysical techniques and imaging methods make it possible to spatially resolve temporal changes at intermediate scales away from boreholes [e.g., *Rubin and Hubbard*, 2005].

Surface-deployed ground penetrating radar (GPR) reflection data [*Talley et al.*, 2005; *Tsoflias and Becker*, 2008; *Becker and Tsoflias*, 2010] has been successful in imaging saline tracers in individual sub-horizontal fractures, but such experiments are limited to depths of some tenths of meters even under ideal conditions. Cross-hole difference-attenuation radar tomography [e.g., *Liu et al.*, 1998; *Day-Lewis et al.*, 2003] can image bulk changes caused by tracer movement through fracture zones at larger depths, but the resolution of the resulting tomograms is insufficient to image the transport in individual mm-aperture fractures. *Lane et al.* [1996] demonstrated significant temporal changes between single-hole GPR reflection data acquired before and after a steady-state saline tracer experiment at Mirror Lake, New Hampshire.

Saline tracer injections increase temporarily the electrical conductivity of the fluid within the fractures contributing to tracer movement and within the pumping boreholes. GPR reflections vary with fluid conductivity, as it affects (1) GPR reflection/transmissivity coefficients at rock-fluid interfaces of fractures and boreholes [e.g., *Tsoflias and Becker*, 2008], and (2) the frequency-dependent attenuation within the fluids [e.g., *Liu et al.,* 1998]. To compare GPR reflections at different observation times, it is necessary to correct for variations in the effective source wavelet that are due to increased fluid conductivity in the observation borehole.

We present single-hole GPR monitoring results acquired during a dipole (injection-extraction) tracer experiment in a well-studied granitic aquifer [*Le Borgne et al.*, 2007]. The main differences with the work of *Lane et al*. [1996] is that (1) we perform a pulse injection to determine the velocity of the tracer front, (2) we account for time-varying electrical conductivity in the observation borehole and associated changes in the ringing noise characteristics, and (3) we perform depth-migrations of the difference data. To the best of our knowledge, this is the first time tracer transport in a network of connected fractures has been imaged using single-hole GPR reflection monitoring. Our results are compared with migrated single-hole GPR reflection sections [*Dorn et al*., in press] acquired under natural flow and tracer-free conditions to determine which of the imaged fractures that transport the saline tracer and how these fractures are inter-connected.

The objectives of this paper are to show that single-hole GPR reflection data combined with saline tracer experiments make it possible to (1) monitor tracer transport in individual fractures, (2) obtain high-resolution depth-migrated images of tracer displacement, and (3) retrieve site-specific information about the geometry and hydrological connections of the main transport pathways.

**2. Field site and experiment**

The tracer test was carried out in a saturated and sparsely fractured (<1 open fracture/m) granite formation close to Ploemeur, France [*Le Borgne et al.,* 2007]. We used two 6 m spaced boreholes B1 (80 m deep) and B2 (100 m deep) that reach a contact zone at $z$ = ~40 m depth ($z$ = 0 m corresponds to the top of the borehole casing) between porous mica schist and underlying low-porosity granite, which is the formation of interest in this study.

Single- and cross-hole GPR reflection data acquired under natural flow conditions [*Dorn et al*., in press] constrain the geometry of the main fractures (midpoint, minimum extent of fracture, fracture dip) within the granite at radial distances $r$ = 2-20 m away from the boreholes. Additionally, borehole logging (optical, acoustic, gamma-ray and resistivity logs) and hydraulic testing characterized fractures that intersect the boreholes and identified those that are hydraulically connected. The formation is highly transmissive with overall transmissivities over the length of each borehole on the order of $10^{-3}$ m$^2$/s. *Le Borgne et al.* [2007] found that the local conductive fracture network is dominated by only a few well-connected fractures (i.e., only 3-5 such fractures intersect a borehole over its entire length), but no single fracture appears to connect B1 and B2.

In the injection well B2, 94 L of saline tracer (50 g NaCl/L, initial tracer salinity is 30 times higher than the background salinity) were injected during 11 minutes at a rate of 8.5 L/min within a transmissive fracture at $z = 55.6$ m that was isolated from lower-lying fractures using a packer system. After the injection, we continued to push the tracer with fresh water at the same rate.

In the observation well B1, we pumped water at ~5.5 L/min and acquired single-hole GPR data (250 MHz omni-directional antennas with a dominant frequency of 140 MHz, 4 m antenna spacing) with a depth sampling of $\Delta z = 0.1$ m over $z = 35$-$75$ m. We used 250 MHz antennas to obtain a high resolution even though *Tsoflias and Becker* [2008] have shown that lower frequency antennas are more sensitive to salinity changes. Thirty GPR raw sections $\mathbf{D}_1^{raw}$ to $\mathbf{D}_{30}^{raw}$ together with borehole fluid electrical conductivity and pressure logs (logger attached to the antenna cables just above the upper antenna) were acquired at observation times $t^{obs}$ relative to the start of the injection. The reference section $\mathbf{D}_1^{raw}$ was acquired just before the injection, and the following sections $\mathbf{D}_i^{raw}$ were acquired every 10 minutes (the acquisition of one GPR section takes approximately 5 minutes), except $\mathbf{D}_{30}^{raw}$ that was acquired the next day after overnight pumping. Relative vertical positioning accuracy of a few cm between time-lapses was possible by using a calibrated digital measuring wheel, and by marking the start and end points on the cables. Two plastic centralizers attached to each antenna assured that the lateral positions within the boreholes were similar between surveys.

## 3. Data processing

Apart from standard GPR processing, we accounted for (1) depth-positioning uncertainties on the cm-scale, (2) temporal variations in the effective source signals caused by variations in the borehole fluid conductivity, and (3) significant direct wave energy and ringing signals caused by poor dielectric coupling that dominate the individual raw sections $\mathbf{D}_1^{raw}$ to $\mathbf{D}_{30}^{raw}$ at traveltimes $t < 90$ ns. Generally, the raw data have high signal-to-noise-ratios for $t < 160$ ns.

Static corrections accounted for time-zero drifts and residual misalignments of the direct wave between individual sections. An initial geometrical scaling was applied together with a wide bandpass filter in the frequency domain (linear tapered with corner frequencies 0-20-300-380 MHz,). We accounted for depth positioning errors by maximizing the correlation between the zero-crossing patterns of individual data traces compared to the stacked section of all data (the standard deviation of the corrections was 3 cm).

To correct for temporal changes of the effective source signal, we applied a continuous wavelet transform and analyzed the wavelet power spectra of the data using the Morlet wavelet [*Torrence and Compo*, 1997]. Firstly, we removed the wavelet scales with corresponding center frequencies outside of the 20-160 MHz range. Secondly, wavelet-scale dependent factors $\mathbf{F}_i$ were defined as the ratios of wavelet power of the direct wave of the processed data $\mathbf{D}_i^{proc}$ with respect to the reference $\mathbf{D}_1^{proc}$ ($=\mathbf{R}_1^{proc}$). We then used the factors $\mathbf{F}_i$ to rescale $\mathbf{R}_1^{proc}$ in the wavelet domain into new reference sections $\mathbf{R}_2^{proc}$ to $\mathbf{R}_{30}^{proc}$. The underlying assumption for this correction is that the increased electrical conductivity of the borehole fluid affects the direct wave in the same way as later arriving signals, such that any remaining differences between time-lapses only reflect changes occurring within the rock formation. We rescaled $\mathbf{R}_1^{proc}$ instead of $\mathbf{D}_i^{proc}$ because of the higher bandwidth of $\mathbf{R}_1^{proc}$ as attenuation of higher frequencies increases towards later acquisition times due to the increasing borehole fluid conductivity. An eigenvector filter applied in a window around the direct wave ($t < 90$ ns) removed ringing effects parallel to the direct wave in the individual sections $\mathbf{R}_i^{proc}$ and $\mathbf{D}_i^{proc}$.

To facilitate amplitude comparisons for different traveltimes and time-lapses we calculated the relative differences $\mathbf{M}_i$ by multiplying the differences $\mathbf{D}_i^{proc} - \mathbf{R}_i^{proc}$ with the inverse envelope sections of $\mathbf{R}_i^{proc}$. We defined a minimum amplitude threshold for the envelope sections of $\mathbf{R}_i^{proc}$ to avoid overestimating difference energy in low-reflectivity regions. The estimated relative difference magnitudes vary smoothly between time-lapses, the main changes occur during the first few time-lapses, and the signal returns towards the background at the end of the experiment (not shown).

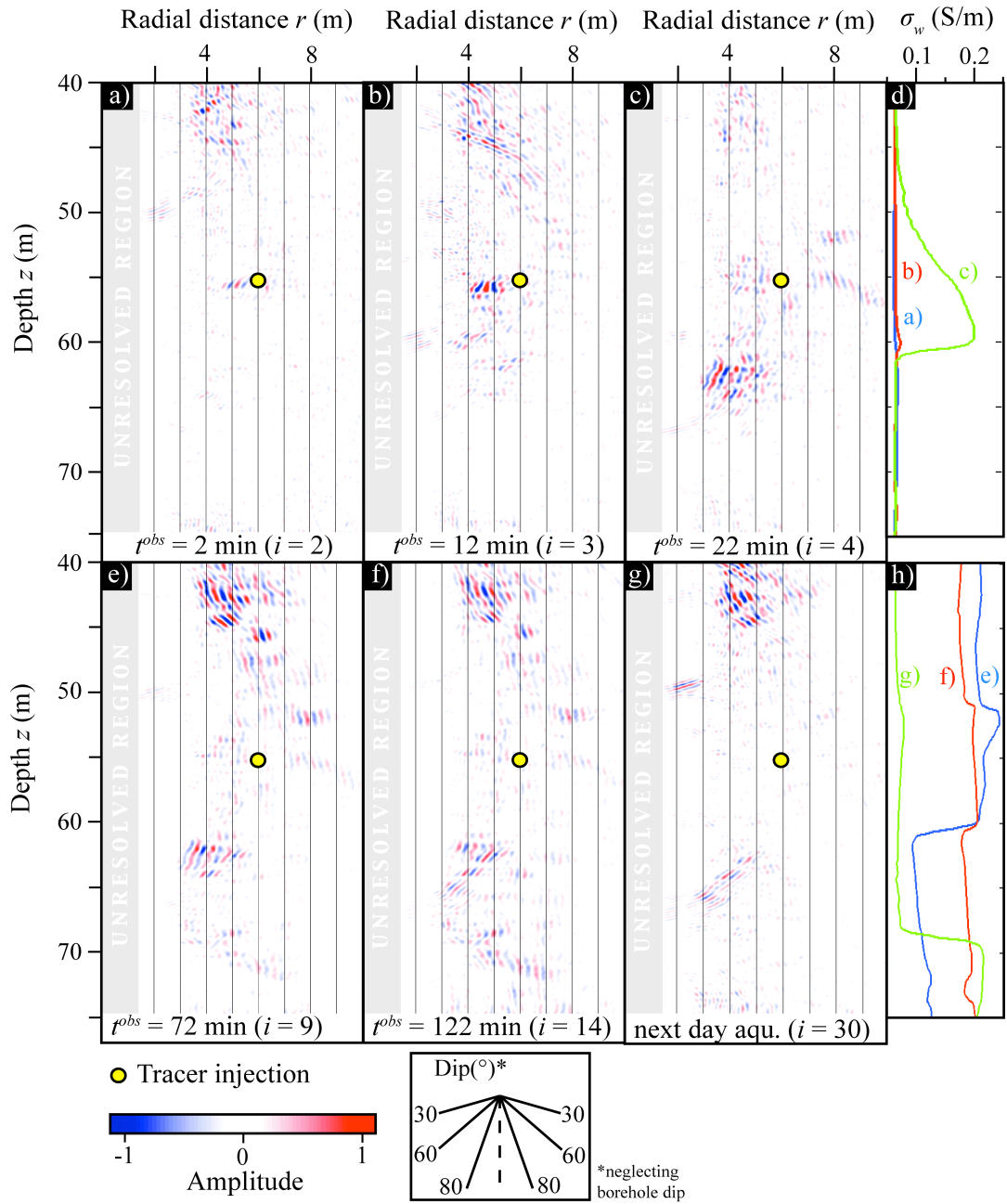

**Figure 1:** Migrated relative difference GPR sections acquired in B1 at $t^{obs}$ (a) 2 min, (b) 12 min, (c) 22 min, (e) 72 min, (f) 122 min, and (g) next day acquisition. (d) and (h) Electrical conductivity $\sigma_w$ of the borehole fluid in B1 acquired at the observation times $t^{obs}$ in (a-c) and (e-g), respectively. High difference patterns originate from increased salinity in fractures located at the front of each such pattern (i.e., smallest radial distance $r$ for each depth $z$). Note that we do not image any features at $r < 1.5$ m (gray region) because of the dominance of the direct wave at early times and its subsequent removal, which tends to remove superimposed reflections at early times.

Pre-stack Kirchhoff depth-migration based on the 1-D velocity function of *Dorn et al.* [in press] made it possible to migrate $\mathbf{M}_i$ with minimal smearing or other artifacts (Figure 1). Migration of difference sections is useful as the linearity of migration in the input wavefield term makes the final migrated sections comparable to GPR results obtained under natural flow and tracer-free conditions [*Dorn et al., in press*]. The unmigrated difference sections $\mathbf{M}_i$ (not shown) contain significant random noise at $t > 130$ ns, but the destructive superposition of random noise energy during migration significantly decreases the presence of incoherent events in the migrated images.

**4. Discussion**

The migrated relative difference sections of $\mathbf{M}_i$ (6 of them are shown in Figure 1) display patterns of high magnitudes with subhorizontal to vertical dips (30-90°, relative to the surface) at $r = 2$-$10$ m radial distance from B1. Note that we only image changes in a 2-D projection around the borehole (i.e., depth $z$ and radial distance $r$). In general, patterns close to the injection point are imaged only at early times $t^{obs}$. Further away from the injection point, patterns appear at later $t^{obs}$ and they are visible for longer time periods.

The evolving magnitudes can be traced from the injection point at $z = 55.6$ m in B2 throughout the depth interval $z = 40$-$72$ m. The dips and locations of these features correlate well with previously imaged fractures from static multi-offset single-hole data (Figure 2a) [*Dorn et al., in press*] and hydrogeological studies (see Figure 2a) [*Le Borgne et al.*, 2007]. This makes us confident that we can identify the main tracer-occupied fractures in Figure 2a by superimposing migrated difference sections on the static images from B1 and in B2 (Figure 2b) by identifying the same fractures as in B1 based on their dips and depths.

At $t^{obs} = 2$ min (Figure 1a), high magnitudes are concentrated in front of the injection interval (C1 in Figure 2) indicating that we can image the injected tracer. Other imaged features correlate with a fracture intersecting B1 at 50.9 m and a complex subvertical fracture zone pattern at $z = 40$-$47$ m (see Figure 2). We think that these changes possibly relate to reactivation of remaining tracer from previously performed saline tracer experiments, performed in the preceding days, as we change the pumping and injection configuration. At $t^{obs} = 12$ min (Figure 1b), we find that the pattern around the injection point has grown in size and magnitude. At the same $t^{obs}$, a feature appears at $z = 60$ m that is related to the fracture dipping 30° (C2 in Figure 2) through which the first saline tracer arrive in B1 at this time (Figure 1d). The downward movement of the tracer continues at $t^{obs} = 22$ min (Figure 1c) with a dip of 75° at $z = 62$-$65$ m (C3 in Figure 2) and a reflectivity pattern appear dipping 50° at $z$=63-66 m (C4 in Figure 2). A pattern at $t^{obs}$ = 22 min and $r > 7$ m appear that is likely related to the feature (Z in Figure 2) with prominent to moderate magnitudes developing at $z$ = 40-57 m (Figure 1e-g). Between $t^{obs}$ = 72-122 min a patchy moderate magnitude pattern develops at $z > 67$ m (C5 in Figure 2) indicating continuous downward movement of the tracer. The next day acquisition ($i$ = 30, Figure 1g) shows a few locations of remaining high magnitudes (C4, Z, and the fracture intersecting borehole at 50.9 m that now reappear), but most of these patterns have disappeared indicating a return to background conditions.

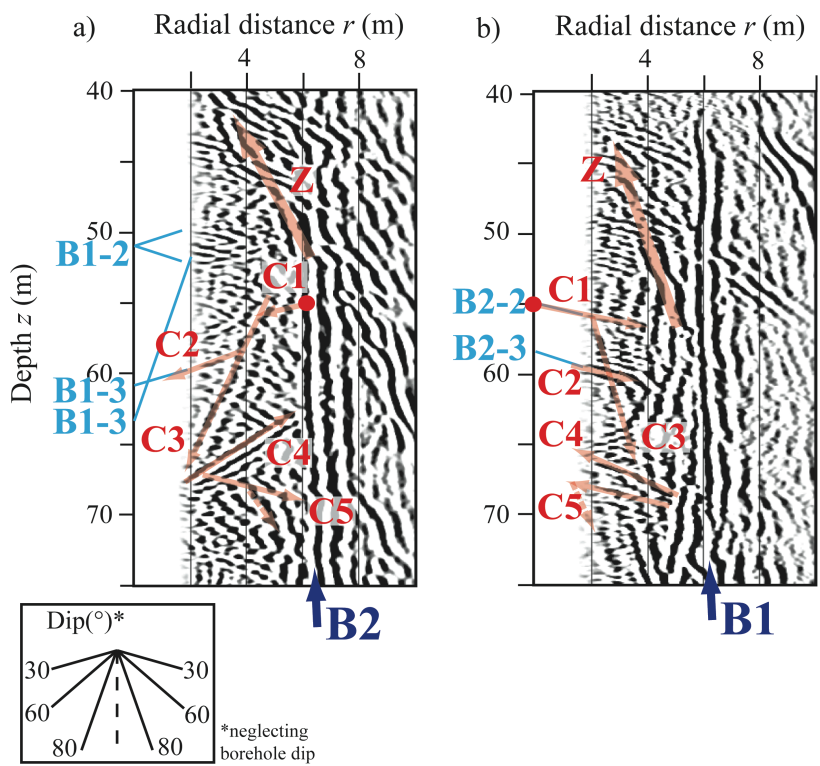

**Figure 2:** Extracts of migrated static single-hole GPR sections of (a) B1 and (b) B2 from *Dorn et al.* [in press] with superimposed interpretation (orange arrows) of fractures (C1-C5, Z) through which the injected tracer moves and the possible flow direction. Light blue letters refer to transmissive fractures identified in the boreholes using optical logs and flowmeter tests [*Le Borgne et al.,* 2007], whereas marine blue letters refer to reflections from other boreholes.

It is possible to follow the tracer between time-lapses, which allows us to estimate approximate velocities of the tracer front (e.g., $v$ = 0.4-0.7 m/min in fractures C2 and C3). The upward movement of the tracer in Z is likely related to the natural upward pressure gradient at the site (i.e., there is an ambient upward flow of ~1.5 L/min in B1) [*Le Borgne et* *al.*, 2007]. The slower changes between time-lapses at later times $t^{obs} > 72$ min may indicate that the remaining tracer is largely unaffected by the pressure gradient imposed in the injection and extraction borehole, and rather follow natural flow gradients (e.g., C4 and Z in Figure 2).

There is significant tracer arrival in the observation borehole B1 through C2 (see Figure 1d and 1h) in accordance with the imaged magnitude patterns at early times. There is also a deeper arrival of saline tracer arriving after more than one hour, which is likely to occur at 78.7 m [*Le Borgne et al.*, 2007] (see Figure 1h). This later and deeper tracer arrival gives confidence in the hydrological significance of C5 (Figure 1e-g). The reflectivity corresponding to the fracture at $z = 50.9$ m for early $t^{obs}$ and the next day acquisition seems uncorrelated with the information from the electrical conductivity logs as this fracture appear to produce fresh groundwater (see Figure 1h). This highlights the need to complement this type of experiment with other data that track the arrival of the saline tracer in the borehole. Furthermore, we argue that televiewer data and flowmeter measurements [*LeBorgne et al.*, 2007] offer an excellent complement to characterize the near-borehole environment as fractures within $r<1.5$ m cannot be imaged with the GPR data.

## 5. Conclusions

We find that single-hole GPR reflection imaging is capable of monitoring saline tracer movement through a connected network of mm-aperture fractures over tens of meters. This was made possible through careful positioning, a rather elaborate processing of the single-hole GPR data and by depth-migrating the relative-difference sections. The processing scheme accounted for variable borehole fluid conductivities, variable transmitter power and geometrical inaccuracies.

The final migrated relative-difference sections image spatially- and temporally evolving patterns at $r = 2$-10 m radial distance with dips between 30-90° at locations corresponding to previously imaged fractures. We conclude that the saline tracer occupies at least 5 fractures (C1-C5) and a large fracture zone (Z) during the tracer experiment. As single-hole GPR reflection data with omni-directional antennas provide 2D projections of reflections, we can only assign a minimum velocity of tracer movements (e.g., ~0.4 m/min for C4). One main advantage of these types of experiments compared to classical tracer experiments is that they provide a length scale of the tracer transport path that can be used together with the breakthrough data to determine transport velocities.

Data from this experiment and another five unreported tracer experiments will in the future be combined with flowmeter data and hydrogeological modeling to constrain the possible geometry and fracture properties of the hydrologically most prominent fractures at the site.

**Acknowledgements**

We thank the field crew of Vincent Boschero, Maria Klepikova, and Nicola Lavenant. We are thankful to Alan Green at ETH-Zurich for making his GPR equipment available. John W. Lane and Frederick Day-Lewis at the USGS provided some very useful advice about the field experiment. Reviews of Matthew W. Becker and an anonymous reviewer helped to improve the manuscript. This research was supported by the Swiss National Science Foundation under grant 200021-124571 and the French National Observatory H+.